\documentclass[12pt]{iopart}

\usepackage{graphicx}
\usepackage{iopams}  
\usepackage{color}

\expandafter\let\csname equation*\endcsname\relax
\expandafter\let\csname endequation*\endcsname\relax
\usepackage{amsmath}

\begin{document}

\title{QED with magnetic textures}

\author{Mar\'ia Jos\'e Mart\'inez-P\'erez and David Zueco}
\address {Instituto de Ciencia de Materiales de
  Arag\'on and Departamento de F\'isica de la Materia Condensada ,
  CSIC-Universidad de Zaragoza, Pedro Cerbuna 12, 50009 Zaragoza,
  Spain}
\address{Fundaci\'on ARAID, Avda. de Ranillas, 50018 Zaragoza, Spain}
\ead{pemar@unizar.es, dzueco@unizar.es}


\begin{abstract}
Coherent exchange between photons and different matter excitations (like qubits, acoustic surface waves or spins) allows for the entanglement of light and matter and provides a toolbox for performing fundamental quantum physics. 
On top of that, coherent exchange is a basic ingredient in the majority of quantum information processors.
In this work, we develop the theory for coupling between magnetic textures  (vortices and skyrmions) stabilized in ferromagnetic nanodiscs and photons generated in a circuit.
In particular, 
we show how to perform broadband spectroscopy of the magnetic textures by sending photons through a transmission line  and recording the transmission.
We also discuss the possibility of reaching the strong coupling regime between these texture excitations and a single photon residing in a cavity.
%
\end{abstract}

\maketitle

\section{Introduction}

With the promise of developing quantum technologies, in the last years an enormous effort has been focused on building different quantum systems operating in a fully quantum coherent way \cite{Acin2018}.
Ion traps, quantum dots and superconducting circuits are prominent examples \cite{Ladd2010}.
Thanks to this progress, it has been realised that combining different physical systems could help in optimising a quantum processor \cite{Xiang2013}.
For example, we can imagine long decoherence spin qubit ensembles working as quantum memories.  This ensemble could be coupled to superconducting transmission lines, where microwave photons can share the information between such a memory and the processing unit made of superconducting  qubits.
On the other way, merging  different substances can serve to transduce between different type of quantum excitations.
Examples of the latter are  mechanical oscillations interacting with superconducting circuits or optical cavities \cite{OConnell2010, Verhagen2012} or coupling surface acoustic waves and superconducting circuits \cite{Gustafsson2014, Manenti2017}.

Another interesting interface would be that formed by spin waves and microwave photons.
This is appealing since photons are well established  in both quantum computing and communication architectures for mediating interactions between qubits and  acting as information carriers, respectively. Spin waves (or their corresponding quasiparticles, magnons) are, on the other hand, their short-wavelength counterpart in spintronics.
Based on their small wavelength, together with the absence of Ohmic dissipation, the field of magnonics aims to exploit spin waves to produce nanoscopic low-loss  devices \cite{Chumak15}.
Different magnonic excitations have been proposed or have already been coupled to quantum light experimentally \cite{Lachance2019}.  
Most studies have focused on Yttrium--Iron--Garnet (YIG) films or spheres coupled to either superconducting coplanar waveguide (CPW) resonators \cite{Huebl13,Morris17} or 3D cavities \cite{SoykalFlatte10,ZahngNpj15}. The latter allows exciting mostly the uniform Kittel mode in which all spins precese in unison. 
%
%
Fewer works have analyzed the case of higher-order spin wave modes in confined geometries \cite{Zhang16,Bourhill16,Osada2018,Cao2015} and individual magnetic solitons such as vortices in soft-magnetic discs \cite{Graf18,MartinezPerez2018}.
%
 Magnetic vortices are extremely stable magnetic textures exhibiting a very rich dynamical  behaviour in the sub-GHz to tens of GHz range. Vortices have been used, \emph{e.g.}, as spin-torque nanoscillators \cite{Pribiag07} or to generate ultrashort ($< 100$ nm) spin waves \cite{Wintz16,Dieterle19}. The latter application is enormously interesting for the implementation of magnonic devices as it would allow the emission of coherent, ultrashort spin waves of arbitrary wavelength.

In a previous work, we have proposed a set up where photons are coupled to the gyrotropic motion of a vortex in a ferromagnetic disc \cite{MartinezPerez2018}.
Importantly, we explored the possibility of reaching the so-called strong coupling regime.  
The latter means that quanta of vortex motion and photons are exchanged in a fully quantum coherent way, because the vortex-photon coupling is larger than both the  material's damping and the photon leakage.   
In the present work, we generalize our proposal to include the coupling to higher-order vortex modes reaching the 15 GHz range.
High-energy modes might lead to larger coupling factors because the intensity of the zero-point current fluctuations in a cavity depends linearly on its resonance frequency. This regime will also be very interesting for vortex-mediated generation of spin-waves that usually takes place within the $1-15$ GHz range \cite{Wintz16,Dieterle19}.

%
We also consider the coupling of microwave photons to the characteristic breathing mode of a magnetic skyrmion confined in a thin magnetic disc. 
 Skyrmions constitute a new paradigm in condensed matter physics. Topological protection makes them enormously stable against thermal fluctuations or material defects  \cite{Bogdanov06}. Even more important for spintronic applications, skyrmions can be moved with record low-power electric currents \cite{Fert13}.

 As we will show here, strong coupling between microwave photons and high-order vortex modes is feasible. Even if the amplitude of these modes is considerably lower than that of the gyrotropic mode, this is compensated by the larger intensity of the electromagnetic fluctuations. Coupling between photons and the breathing mode peculiar to skyrmions is much more tricky. All these couplings are ultimately limited by the Gilbert damping of the ferromagnetic material.
 We will start by describing the spin Hamiltonian and the resulting stable magnetic textures and normal modes.
 The rest of the paper is organized as follows.
 We introduce the theory for coupling magnetic textures to photons generated in superconducting circuits both in open transmission lines and single mode cavities. 
 We review the different vortex modes and how to measure it by means of a transmission experiment.
 Finally, we explore the coupling regimes of both vortices and skyrmion modes to single cavity photons and discuss the feasibility of reaching the strong coupling regime.

\section{Magnetic Hamiltonian and its coupling to light}

\subsection{Magnetic textures}

The  spin Hamiltonian of a ferromagnet can be written as:
\begin{equation}
\label{HS}
   H_S=H_{\rm ex} + H_{\rm dip} +H_{\rm K} + H_{\rm DMI} +H_{\rm Z}.
\end{equation}
Here, $H_{\rm ex} = - J_{ij} \vec S_i \, \vec S_j $ is the Heisenberg exchange energy,  $H_{\rm dip}$ is the dipolar term that acounts for the total magnetostatic energy of the system  (shape anisotropy), $H_{\rm K} $ is the magnetocrystalline anisotropy that sets some preferred axis for the magnetization and $H_{\rm DMI} = D_{ij} \vec S_i \times \vec S_j$ is the Dzyaloshinskii--Moriya interaction (DMI) term, with $D_{ij}$ the strength of the asymmetric interaction. Finally, $H_{\rm Z}=  - g_{\rm e} \mu_B \vec B ({\bf r}_i) \, \vec S_i$ is the Zeeman coupling between spins and the externally applied magnetic field  $\vec B ({\bf r}_i)$.
$\vec { S}_i = (S_i^x, S_i^y, S_i^z)^t $ are spin-angular momentum operators, $g_{\rm e}$ is the electron g-factor and $\mu_{\rm B}$ is the Bohr magneton.
The spin operators satisfy  $[S^\alpha_i, S^\beta_j]= {\mathrm i} \delta_{ij} \, \epsilon_{\alpha \beta \gamma} S^\gamma$ with $\alpha, \beta, \gamma = x,y,z$.
%
In this paper, the  characteristic length scale of the magnetic textures under study and their excitations are much  smaller than the dimensions of the nanodiscs where they are stabilized. 
Under these conditions, quantum fluctuations can be neglected and the equations of motion for the average magnetization, \emph{i.e.}, $\vec m_j = g_{\rm e} \mu_B \langle \vec S_j \rangle$, can be casted in the Landau-Lifshitz-Gilbert (LLG) equation \cite{Landau,Gilbert}:
\begin{equation}
   \frac{\partial \vec m_j}{\partial t} = \frac{\gamma_{e}}{1+\alpha_{\rm LLG}^2} \Big\{
   \vec m_j \times \vec B_{\rm eff} + \alpha_{\rm LLG} \big[ \vec m_j \times (\vec m_j \times \vec B_{\rm eff} )   \big]
   \Big\}.
    \label{LLG}
\end{equation}
Here, $\vec m_j = (m^x_j, m^y_j, m^z_j)$,  $\gamma_{e}$ is the electron gyromagnetic ratio, $\alpha_{\rm LLG}$ is the dimensionless Gilbert damping of the material, and  $\vec B_{\rm eff} $ is an effective field that accounts for all terms considered in the spin Hamiltonian \eqref{HS}.

\begin{figure}
\begin{center}
\includegraphics[width=4in]{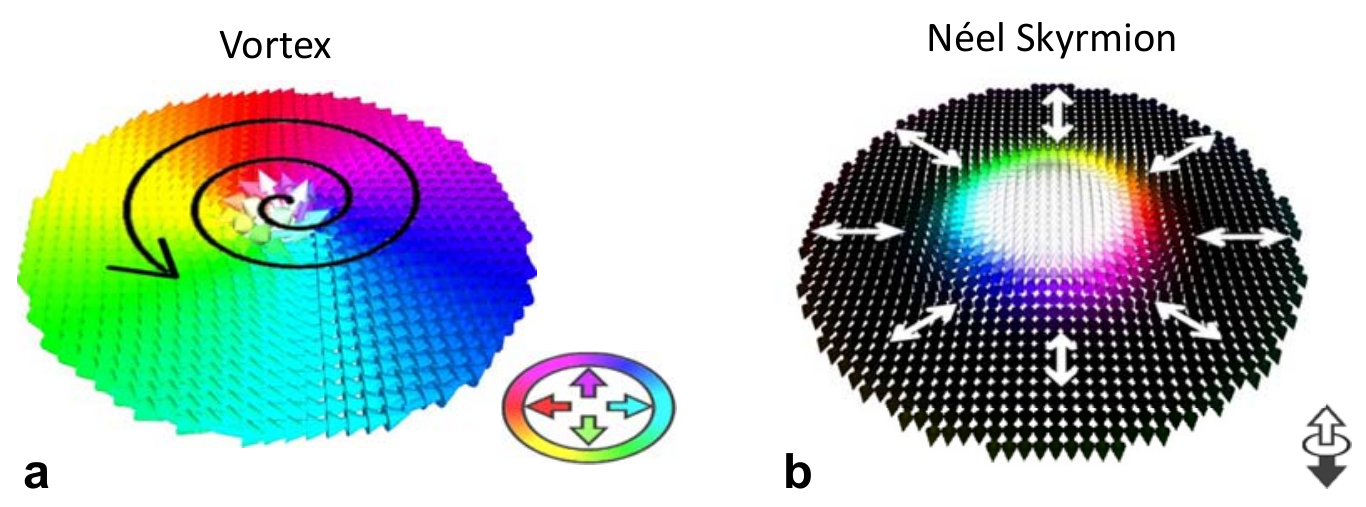}
\end{center}
\caption{\label{fig:sketch} Spatial distribution of magnetic moments typical of  (a) a vortex  or (b)  a N\'eel skyrmion confined in thin magnetic discs. In-plane magnetic moments are coded according to the colour wheel legend on the bottom right part of panel (a), whereas white/black out-of-plane moments point up/down according to the legend in the bottom rigth part of panel (b). Black arrows indicate the magnetization motion corresponding to the lowest energy modes, \emph{i.e.}, gyrotropic and breathing, for the vortex and the skyrmion, respectively.  
}
\end{figure}

Depending on the magnetic object under study (material and shape) and the initial conditions (or the magnetic history), relaxing Eq. \eqref{LLG} leads to the stabilization of different magnetic textures. These might have very interesting properties such as topological protection which makes them extremely stable against, \emph{e.g.}, thermal fluctuations or material defects. Two archetypal topologically protected textures are the magnetic vortex  and the skyrmion. The former is the ground magnetic state in flat micro- or nanoscopic magnetic discs with negligible magnetocrystalline anisotropy [\emph{i.e.}, $H_{\rm K}=0$ and $H_{\rm DMI}=0$ in the spin Hamiltonian \eqref{HS}]. The minimization of surface magnetic charges makes spins lie preferentially parallel to the borders of the disc, leading to the characteristic vortex spin curling in clockwise or counterclockwise fashion, as shown in Fig. \ref{fig:sketch}a. In the vortex core, spins turn out-of-plane pointing up or down defining the vortex polarity \cite{Shinjo00}. This yields four possible states that are, in principle, degenerate. The lowest energy excitation peculiar to vortices is the gyrotropic mode, in which the vortex core gyrates around its equilibrium position. It is predominantly excited by means of an in-plane excitation magnetic field.

Skyrmion stabilization is a little more tricky, since it requires the presence of terms favouring non-collinear spin ordering such as the DMI [\emph{i.e.}, $H_{\rm DMI}\neq 0$ in the spin Hamiltonian \eqref{HS}]. Skyrmions are characterized by a central core pointing in the opposite direction to the surrounding magnetization, so that spins can be projected once onto the unit sphere. N\'eel-like skyrmions can be typically stabilized in confined geometries such as flat micro- or nanodiscs by means of the interfacial-DMI (see Fig. \ref{fig:sketch}b). This interaction arises in multilayers where ultra-thin ferromagnets are combined with materials having large spin-orbit coupling \cite{Finocchio2016}. Additionally, a preferred out-of-plane magnetic ordering is required. The latter is usually achieved by the application of external magnetic fields, or by using materials with perpendicular magnetic anisotropy [\emph{i.e.}, $H_{\rm K}\neq 0$ in the spin Hamiltonian \eqref{HS}]. Using perpendicular (out-of-plane) fields it is possible to excite the breathing mode peculiar to the skyrmion, in which it conserves radial symmetry \cite{Joo-Von14}.


\subsection{Light-matter  Hamiltonian}
\label{sect:general}

The objective of this work is to study the coupling between vortex or skyrmion excitations and photons.
In particular, photons generated by currents in superconducting circuits, either propagating in transmission lines or stationary photons living in cavities.
%
%
The total Hamiltonian, including the magnetic nanodisc, the electromagnetic field and their interaction is decomposed as:
\begin{equation}
\label{HT}
    H_T
    = H_S + H_Q + H_I
    \; .
\end{equation}
Here, $H_S$ is the spin Hamiltonian [cf. Eq. \eqref{HS}] and
$H_Q$ is the photonic Hamiltonian. In the case of one dimensional (1-D) transmission lines it reads \cite{Gu2017};
\begin{equation}
\label{HQ}
    H_Q = 
    \hbar
    \int  d k 
    \;
    \omega_k
    \,
    a_k^\dagger a_k,
\end{equation}
with $[a_k, a_{k^\prime}^\dagger]=\delta(k -k^\prime)$.
Here, we are considering 1-D fields  where the $k$ index stands for the photon wavenumber for the transverse modes solutions of the wave equations.  
Owing to the fact that light propagates in 1-D ($z$ with our choice, cf. Fig. \ref{fig:sketch}), we know that $\omega_k = c k$  with $c$ the propagation velocity. In transmission lines $c= \sqrt{1/LC}$ with $L$ ($C$) being the effective inductance (capacitance) per unit of length.
$H_I$ is the Zeeman coupling between the magnetic moments and the magnetic field $\vec B ({\bf r})$ generated by the circuit:
\begin{equation}
\label{HI}
    H_I = - g_{\rm e} \mu_{\rm B}  
    \sum_n 
    \vec {\mathcal S}_i 
    \,
    \vec B ({\bf r}_i) 
    \; .
\end{equation}
Finally, the magnetic field at point ${\bf r} = (x,y,z)$ is quantized in the Coulomb gauge as \cite [Chapter 10] {Schleich2001}:
\begin{equation}
\label{B}
    \vec B ({\bf r}) = 
    \sqrt{\frac{\hbar}{2  c ^2 \epsilon_0}}
    \begin{pmatrix}
    v_x (x,y)
    \\
    v_y (x,y)
    \\
    0
    \end{pmatrix}
    \int  d k 
    \sqrt{\omega_k}
    \;
    a_k^\dagger e^{i k z} + {\rm h.c.}
\end{equation}
Here, ${\rm h.c.}$ means hermitean conjugate, $v_{x}$ ($v_{y}$) are the transversal $x$ ($y$) components of the field density having units of inverse length, $c$ is the speed of light in vacuum and $\epsilon_0$ is the vacuum permittivity.
We emphasise that we are mainly interested in discussing the single photon coupling, thus these components must be understood as the \emph {root mean square} (rms) value of the  field  generated by the vacuum current in the circuit. 
In section \ref{sect:cQED} we explain how to compute them.
%

\subsection{Waveguide QED and textures normal modes}
\label{sect:cwQED}

%
Quantum systems coupled to 1-D bosonic fields, as in Eqs. \eqref{HI} and \eqref{B}, {\color{red}} are of high interest. They  are named  waveguide quantum electrodynamics (waveguide-QED)  to differentiate them from cavity-QED setups, where the quantum system is coupled to the discrete stationary modes of a cavity.
1-D waveguides are used to enhance the light and matter coupling in order to, \emph{e.g.}, mediate effective interactions between spatially separated quantum systems or to generate or manipulate quantum states of light as single or $N$-photon wavepackets.  Here,  we use this waveguide QED setup to perform broadband spectroscopy of magnetic textures. 

\begin{figure}
\begin{center}
\includegraphics[width=\textwidth]{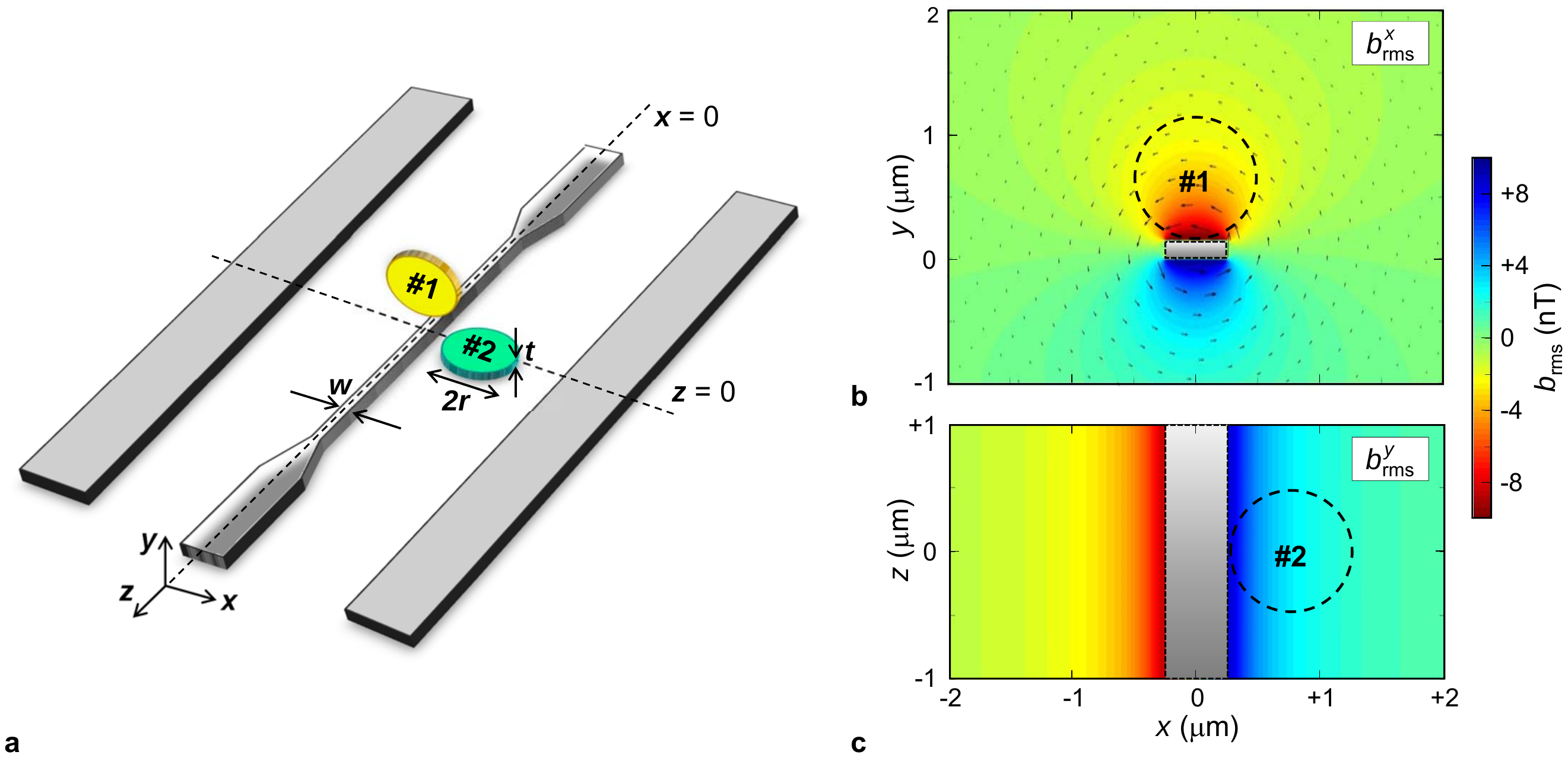}
\end{center}
\caption{\label{fig:fields} (a): Scheme of the proposed experiment. The relevant dimensions and the coordinate axes are highlighted. 
Two different disc locations are considered, $\#1$ and $\#2$. These correspond to the rms excitation field applied in-plane and out-of-plane of the disc, respectively. $2r$ is the disc diameter, $t$ the thickness, $w$ the constriction width. (c) and (d) show the numerically calculated density plots of the $x$ and $y$ components of the rms excitation field along the discs $\#1$ and $\#2$, \emph{i.e.}, planes $z=0$ and $y=0$, respectively. In the calculations we assume a zero-point current $i_{\rm rms} = 11$ nA flowing through a superconducting constriction with $w=500$ nm and thickness $150$ nm. }
\end{figure}

An ideal experimental setup for resolving the excitation spectrum will use
the field fluctuations in a waveguide as the perturbation field, as sketched in Fig. \ref{fig:fields}.
Owing to the continuum spectrum of the waveguide (using superconducting circuits, the operation range can be safely assumed to lie between few MHz up to 15-20 GHz) we can access all characteristic modes in a single experiment.
In particular, the excitation spectrum can be obtained  by measuring the transmission  through the superconducting waveguide.
%
%
To see how it works, we study the dynamics of a nanodisc coupled to a waveguide through Eq \eqref{HI}.
We highlight that we will consider two locations, \emph{i.e.}, position \#1 and position \#2 in Fig. \ref{fig:fields}a. The former will be used to excite modes susceptible to in-plane magnetic fields. Under these circumstances, the rms  field is negligible along the disc thickness ($z$ direction, $b_{\rm rms}^z \sim 0$) but it is non-homogeneous along the $(x, y)$ plane as it can be seen in Fig. \ref{fig:fields}b. On the other hand, position \#2 will be used to excite modes susceptible to an out-of-plane excitation. In this case, $b_{\rm rms}^x \sim 0$ and the non-homogeneous $b_{\rm rms}^{y,z}$ is plotted in Fig. \ref{fig:fields}c.
%
%
Conveniently, the excitation magnetic field can be substituted by  the one passing through the center (core) of the topological solution.
The latter is a good approximation for the gyrotropic, first azimuthal modes and the breathing mode peculiar to skyrmions since the magnetization modulation concentrates mostly on the disc's central region. This is not the case of higher-order azimuthal and radial modes.  In these cases, however, our approach underestimates the resulting modulation of the magnetization by a factor $\sim 1.5$ only, as checked numerically.
Thus, we can approximate the relevant field $\vec B ({\bf r}) \cong \vec B ({\bf r}_c)$, with ${\bf r}_c$ the center of the nanodisc.  In this way, $\vec B ({\bf r}_c) = B^x ({\bf r}_c)$ at position \#1 whereas $\vec B ({\bf r}_c) = B^y ({\bf r}_c)$ at position \#2.
Consequently, we can write a simplified  total coupling Hamiltonian
\begin{equation}
\label{HwQED}
  H_T
  =
  H_S
  +
  \int d k  \, \omega_k a_k^\dagger   a_k
  +
  \sum_{\alpha=x,y} \sum_j S^\alpha_j \, \int  \lambda_\alpha(\omega_k) \;  e^{i \omega z /c} a_k^\dagger + {\rm h.c.},
\end{equation}
with
$
\lambda_\alpha (\omega_k) =  \sqrt{\frac{\hbar}{2  c ^2 \epsilon_0 }} v_\alpha (x_c,y_c)  \sqrt{\omega_k} $
and $\alpha = x$ and $y$ for positions \#1 and \#2, respectively. 
%
%

\begin{figure}
\begin{center}
\includegraphics[width=\textwidth]{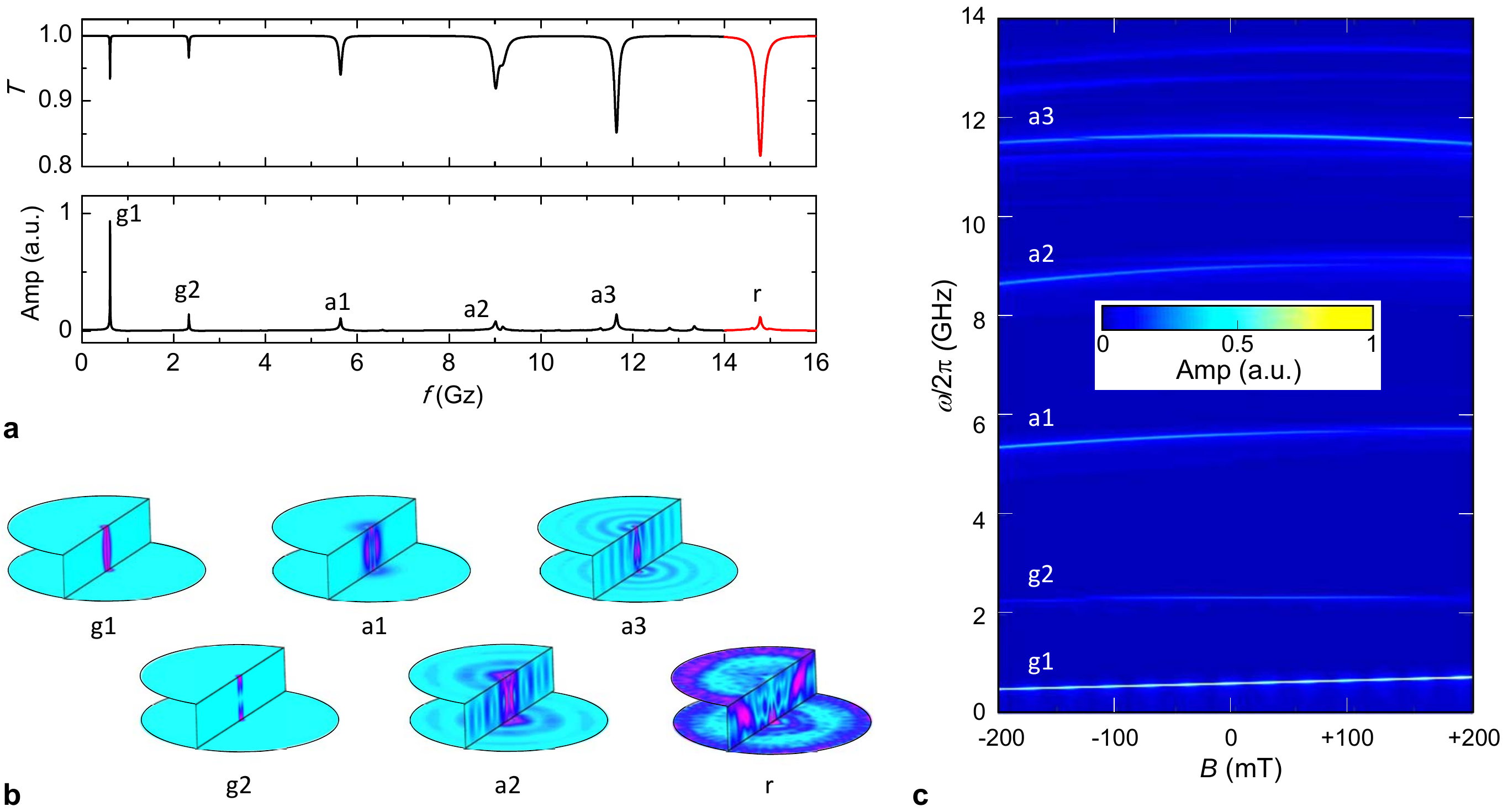}
\end{center}
\caption{\label{fig:modes} Resonant modes of a typical magnetic disc with $r=500$ nm and $t=80$ nm. We use the following notation: 'g' stands for gyrotropic, 'a' denotes the azimuthal modes whereas 'r' refers to the radial mode. (a): Numerically calculated transmission and FFT of the magnetisation response to an excitation magnetic field. The black (red) curve is the response to an in-plane (out-of-plane) oscillating magnetic field. (b): Spatially resolved FFT of the top/bottom surfaces of the disc and a transverse cut corresponding to the different modes. (c): Dependence of the resonant frequencies on the application of a DC out-of-plane magnetic field. Only modes excited by an in-plane excitation field are shown.}
\end{figure}

Using input-output theory it is possible to find the scattering matrix giving for the transmission function 
\cite{Roy2017}:
\begin{equation}
\label{tw}
    T(\omega)= 
    1 - 
    \sum_i \frac{2 \pi \lambda^2(\omega_n)}{2 \pi \lambda^2 (\omega_n) +  \Delta \omega_n + {\mathrm i} (\omega_n - \omega)},
\end{equation}
while the reflection is given by $R(\omega) = T(\omega) -1$. 
The transmitted (reflected) power measured is given by $|T(\omega)|^2$ ($|R(\omega)|^2$).  
Here, $\omega_n$ are the resonant frequencies of the different normal modes whereas $\Delta\omega_n$ are their corresponding linewidths (being proportional to $\alpha_{\rm LLG}$). $\lambda(\omega)$ is the coupling density to the transmission line   that will be calculated explicitly in next section.  Additionally, $n=$ \{g1, g2, a1, a2, a3 and r\} for the gyrotropic, azimuthal and radial modes.

 Numerically calculated values of $|T(\omega)|$ are shown in Fig. \ref{fig:modes}a (upper panel). Simulations are performed  using the GPU-accelerated code MuMax$^3$ \cite{mumax}. We assume a prototype ferromagentic disc with radius $r=500$ nm and thickness $t=80$ nm, saturation magnetization $M_{\rm s}= 1$ MA/m,  exchange stiffness constant $A_{\rm ex}=15$ pJ/m and $\alpha_{\rm LLG} = 10^{-3}$.
 %
Its excitation spectrum is obtained by applying a perturvative field $b(\tau) = A \, {\rm sinc} (2\pi f_{\rm cutoff} \tau)$ with
$f_{\rm cutoff}  = 50$ GHz.
%
Calculating the  Fast Fourier Transform (FFT) of the resulting spatially-averaged time-varying magnetization along the field direction results in the spectrum plotted in Fig. \ref{fig:modes}a (bottom panel). The black curve results when applying an in-plane driving field whereas the red one corresponds to an out-of-plane excitation. Fitting each resonance peak to a Lorentzian function allows estimating $\omega_n$ and $\Delta\omega_n$. These values, together with the estimated $\lambda(\omega)$ computed in next section, are plugged into Eq. \eqref{tw} providing the calculated transmission curve.

In order to have a clearer picture of the magnetization profile of each mode, we plot the corresponding amplitude of the spatial FFT along the upper/bottom disc surface and the transverse cut (see Fig. \ref{fig:modes}b).
The lowest energy mode (g1) corresponds to the vortex core translation around the central position.
The gyration sense is only given by the vortex polarity regardless the direction of the in-plane magnetization which plays no role.
This is the archetypal gyrotropic mode where the vortex core is only minimally distorted through the disc thickness.
At slightly larger energy, we observe the second gyrotropic mode (g2) where the vortex gyrates in the upper/bottom disc surface with a $\pi$ phase shift.
In this way, the vortex core itself flexes through the disc thickness yielding one node in the center.
Several azimuthal modes (a1, a2 and a3) can be observed as well. Here, the vortex core also gyrates but there is an additional magnetization spiral breaking radial symmetry. Magnetization exhibits also a increasing number of nodes along the radial direction for increasing mode number. Importantly, these modes are not homogeneous trough the disc thickness but curl at opposite directions from top to the bottom surfaces \cite{Verba16}.
Finally, in the case of an out-of-plane excitation field one finds a dominant mode conserving radial symmetry (r).
We highlight the importance of performing three dimensional simulations in order to take the disc thickness into account. This is especially true when assuming very thick ferromagnetic discs, like the one studied here.

The energy of each mode can be slightly tuned by means of  an external DC homogeneous magnetic field.
This can be seen in Fig. \ref{fig:modes}c, where the energy spectrum is plotted against a varying out-of-plane homogeneous field. 
The gyrotropic mode increases linearly with the applied magnetic field whereas higher energy modes exhibit a non-monotonous behaviour. 
The presence of other low-amplitude resonant modes can be appreciated as well. 
This is the case of, \emph{e.g}, modes visible between 12-14 GHz that will not be studied here.


\section{Magnetic textures and cavity QED: strong and weak coupling regimes}
\label{sect:cQED}

Shunting the transmission line by two capacitors, it is possible to confine the electromagnetic field. These capacitors act as mirrors creating a microwave realisation of a Fabry-P\'erot cavity.
The boundary conditions at the capacitors impose the frequency quantization for the photonic modes inside the cavity, namely $\omega_n =  c \pi / l  n$, with $l$ the cavity length (the distance between the capacitors).
Then, the nanodisc-resonator can be cast in a cavity-QED-like Hamiltonian [cf. Eq. \eqref{HwQED}]:
\begin{equation}
\label{HQED}
    H_T = H_S + \hbar \omega_C a^\dagger a  +  \sum_{\alpha=x,y} \sum_j S^\alpha_j \,  \lambda_\alpha    ( a^\dagger + a ) 
\end{equation}
where we have applied the single-mode approximation.  The coupling to the cavity mode can be related to the coupling density $\lambda(\omega)$ in \eqref{HwQED} as 
$\lambda_\alpha (\omega ) = \sqrt{\pi} \lambda_\alpha^2/\omega_C$ \cite{Hummer2013}.

Without intrinsic spin-spin interactions,  Eq. \eqref{HQED} reduces to the  Dicke model \cite{Hepp1973}.  
In our case, interactions can not be neglected and solving the full problem is a formidable task.
However, considering that the cavity is fed by only a few photons, ultimately only one, it is safe to assume that the topological solution is perturbed within the linear response regime.
In such a case, the dynamics of the magnetic texture can be cast in a harmonic-oscillator-like equation of motion \cite{Guslienko2006, Krger2007}.
Therefore, the magnetization $M_\alpha= \sum_j m_j^\alpha / V$  with $V$ the nanodisc volume can be quantized as
\begin{equation}
    M_\alpha = \Delta M_\alpha (b^\dagger + b).
\end{equation}
Here, $\Delta M_\alpha = \sqrt{ \langle \Delta M^2 \rangle - \langle M^2 \rangle }$, where $\langle ... \rangle$ are averages over the ground state.   
Analogously, we can rewrite  $\lambda = \sqrt{ \langle \Delta B^2 \rangle - \langle B^2 \rangle} \equiv b_{\rm rms}$. In the last equivalence, we take the usual notation in the literature.
Putting all together, 
the texture excitations coupled to a single mode cavity can be written in it simplest form as two coupled harmonic modes
\begin{equation}
\label{Hharmonic}
    H_T \cong  \hbar \omega_C a^\dagger a +  \hbar \omega_T b^\dagger b +  \hbar g (b^\dagger + b) (a^\dagger + a) \, .
\end{equation}
Here, the coupling strength:
\begin{equation}
\label{Gv0}
    \hbar g = V b_{\rm rms} \Delta M \; ,
\end{equation}
quantifies the exchange rate between single photons and single quanta of texture motion.
The most fundamental figure of merit in cavity QED is the ratio between $g$ and the dissipative rates, namely the  resonance linewidth ($\Delta \omega_n$)  which is proportional to the material damping $\alpha_{\rm LLG}$ and the cavity leakage ($\kappa$).  
If the coherent coupling $g$ exceeds  both dampings one lies within the strong coupling regime \cite{Haroche2013}. In this way, it is possible to  entangle photons and resonant magnetic modes. This serves  to use the cavity photons as a quantum bus among two or more nanodiscs, or to perform a transducer between microwave photons and spin waves.
To estimate if this regime is reachable, the spin excitations and photons must exchange populations coherently in the form of vacuum Rabi oscillations before they are damped out. The condition to have such oscillations is given by:
 $   4 g > | \kappa  - \Delta \omega_n| $ \cite{Auffves2010, Zueco2019}.
%
In our case, $\kappa$ can reach the Hz range easily, whereas typical ferromagnetic materials exhibit $\Delta \omega_n \sim$ MHz at best.  
Therefore, throughout this work, we are interested in checking if 
\begin{equation}
    4 g > \Delta \omega_n
    \label{condition}
\end{equation}
is possible.

To obtain the coupling $\lambda_\alpha (\omega)$ and $g$ we proceed as follows.
We fix our attention to coplanar waveguide architectures, where a central superconducting line of a few microns is  surround by two ground planes.
First, we need to compute the rms of the zero point current fluctuations flowing through the central conductor:  \begin{equation}
    i_{\rm rms} =  \omega_{\rm C} \sqrt{ \frac{\hbar \pi}{2 Z_0} },
    \label{irms}
\end{equation}
with  $Z_0$  the impedance of the resonator. 
The spatial distribution of $i_{\rm rms}$ is calculated using the software 3D-MLSI that allows one to solve the London equations in thin film superconducting wires. 
This current generates a field $\vec B ({\bf r})$ that is used to compute the $b_{\rm rms}$ at the center of the nanodisc.  
Besides, $\vec B ({\bf r})$ also serves to excite the topological solution, yielding $\Delta M$ as  the maximum  response in magnetization.
The coupling is finally calculated using Eq. \eqref{Gv0} and considering the rms magnetic field at the disc's center, \emph{i.e.}, $B^x ({\bf r}_c)$ for position \#1 or $B^y ({\bf r}_c)$ for position \#2 (see Fig. \ref{fig:fields}).

\begin{figure}
\begin{center}
\includegraphics[width=4in]{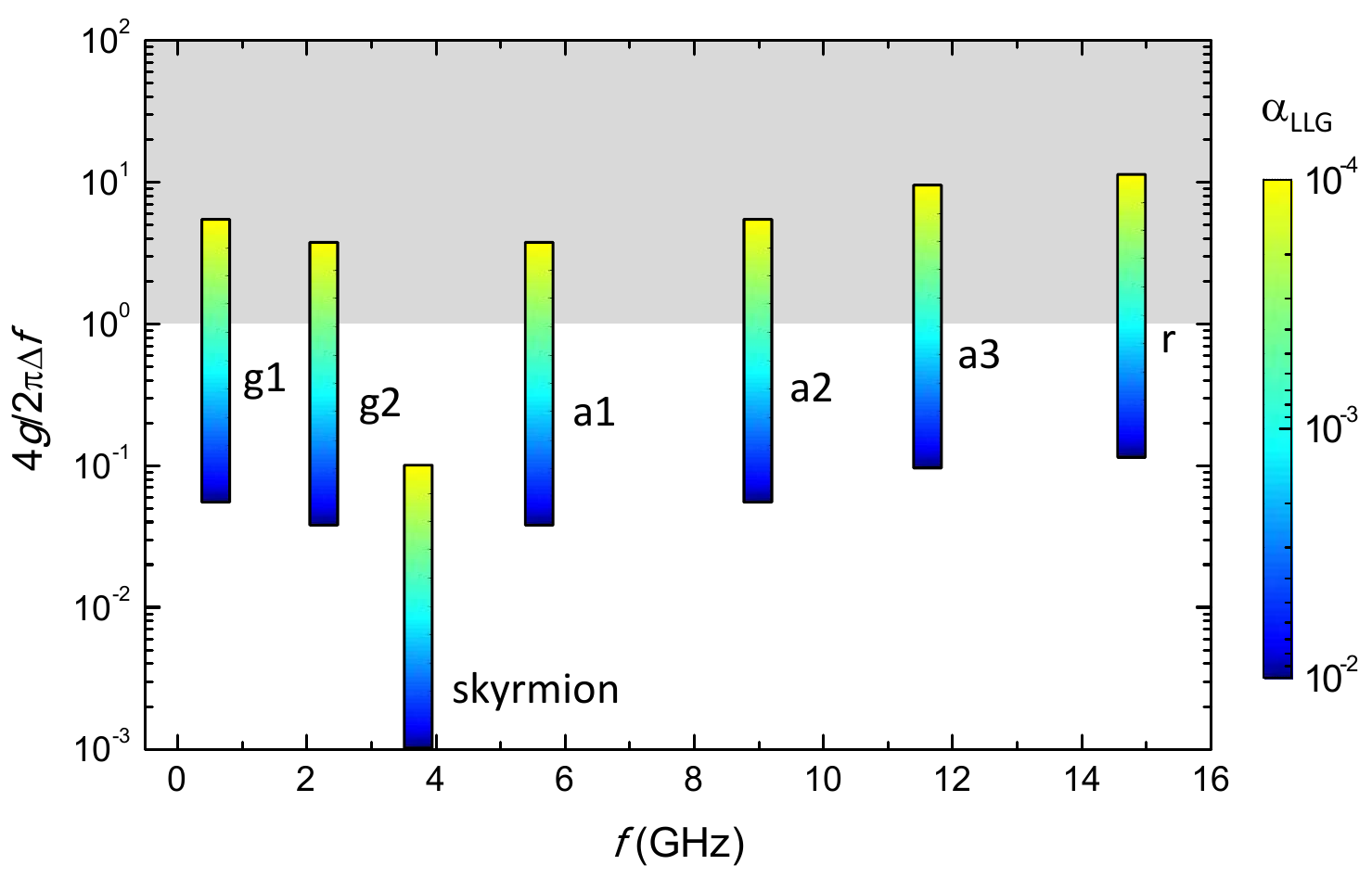}
\end{center}
\caption{\label{fig:coupling} Numerically calculated coupling condition [cf. Eq. \eqref{condition}] for the different resonant modes of the vortex and skyrmion textures. The strong coupling condition increases linearly with decreasing $\alpha_{\rm LLG}$. The shadowed (white) region corresponds to strong (weak) coupling. These serve to highlight the possibility of reaching strong coupling for all resonant modes of the vortex texture provided the damping is low enough. }
\end{figure}

Regarding the vortex modes, we will assume the same  parameters for the ferromagnetic nanodisc used in the previous section.
In the case of the skyrmion, we assume a thin multilayer disc with $r=50$ nm and $t=1$ nm. Given a prototype perpendicular magnetic anisotropy  constant $K_u=1$ MJ/m$^{3}$ and interfacial DMI strength $D_i =3$ mJ/m$^2$, a N\'eel-like skyrmion will be stabilized in the absence of externally applied magnetic fields.
We highlight that coupling to each normal mode is calculated assuming a CPW resonator with characteristic frequency $\omega_{C} = \omega_{n}$, with $n=$ \{g1, g2, a1, a2, a3, r and s\}.
In addition, we have assumed that a  constriction of width $w$ is patterned in the central conductor to increase the strength of the resulting $b_{\rm rms}$ field. The reader is referred to Fig. \ref{fig:fields} for the definition of $w$ and to references \cite{Jenkins2014} and \cite{MartinezPerez2018} for a discussion on the increase of the coupling strength using nanoconstrictions.
In our calculations, $w=500$ nm  (in a 150 nm-thick superconducting central conductor) for the magnetic vortex whereas $w=50$ nm  (in a 50 nm-thick conductor) for the skyrmion.
Finally, we have considered a low-impedance ($Z_0 = 1$ $\Omega$) cavity to further enhance the zero-point field fluctuations $b_{\rm rms}$ \cite{Eichler17}, cf. Eq. \eqref{irms}.

In Fig. \ref{fig:coupling} we plot the resulting strong coupling condition given by Eq. \eqref{condition} for the different resonant modes and for different values of the damping parameter. 
All vortex modes exhibit similar values of the coupling strength whereas that of the breathing mode peculiar to skyrmions lies almost two orders of magnitude below.
This is mainly due to the different sizes of the discs considered in the simulations.
Ultra-thin films are required to obtain sizeable perpendicular anisotropy constants and interfacial DMI, limiting enormously the maximum thickness of the disc with skyrmionic ground states.
Increasing the saturation magnetization would also yield increased $g$ values but $M_{\rm s}$ is normally limited to values below $1 -2$ MA/m.
For the disc's sizes analyzed here, using materials with low $\alpha_{\rm LLG} \sim 10^3$ will assure reaching the strong coupling regime in the case of vortex resonant modes.
%
%
Typical ferromagnetic metals such as Fe and Py exhibit values of the order of $2 \times 10^{-3}$ and $8 \times 10^{-3}$, respectively \cite{Gladii2017}. 
Much lower values have been reported for Heusler and Co$_x$Fe$_{1−x}$ alloys reaching $10^{-3}$ and $5 \times 10^{-4}$, respectively \cite{Drrenfeld2015,Schoen16}. 
Currently, record low damping values are reported for the insulating ferrimagnet YIG, having $\alpha_{\rm LLG} \sim 5 \times 10^{-5}$, although at the cost of a largely reduced saturation magnetization \cite{Yu2014}.
Reaching strong coupling with the skyrmion is much more difficult.
This would require increasing the intensity of the zero-point field fluctuations in the cavity and using a thin-film magnet exhibiting ultra low damping characteristics together with large perpendicular magnetic anisotropy. 
In this kind of materials, the damping usually lies in  the $\alpha_{\rm LLG} \sim 0.1$ range, although promising values of $ \sim 3 \times 10^{-4} $ have been recently reported in bismuth doped YIG \cite{Soumah2018}.

\section{Conclusions}

Starting from the light-matter Hamiltonian, we have calculated the coupling of vortex and skyrmion modes to both propagating and cavity photons.  Assuming typical material parameters, we have shown that the different modes can be distinguished in a transmission experiment using superconducting transmission lines. Importantly, we have also demonstrated that strong coupling between the different vortex modes in a nanodics  and a single photon in a superconducting resonator is feasible within current technology. For this purpose, low-damping materials such as Heusler and Co$_x$Fe$_{1−x}$ alloys are paramount.
On the other hand, reaching the strong coupling regime with skyrmion modes is much more challenging.
This is due to the conditions imposed on the materials and the thickness of the films where skyrmionic states can be stabilized. The latter requires the use of ultra-thin ferromagnets that come along with high dampings and small thickness. 
To summarize, we have generalized our previous theory [cf. Ref. \cite{MartinezPerez2018}], that was developed  for the gyrotropic mode peculiar to vortices and single cavity photons, to provide a unified picture. In this way, we can treat now the coupling between whatsoever magnetic resonant mode and propagating or cavity photons.

The coupling between resonant modes occurring in vortex or skyrmion states might be important to transduce microwave photons to spin waves and vice versa. The latter are appealing due to their short wavelength that would enable to manipulate information in low-loss nanoscopic devices. Additionally, coherent coupling between different magnetic textures located in different regions of a CPW cavity would be feasible. This would allow phase-locking distant magnetic nanoscillators. Finally, photons could also be used to mediate the coupling between magnetic excitations and superconducting qubits located in the cavity \cite{Lachance2019}.

\section*{Acknowledgements}

 We acknowledge support by the Spanish Ministerio de Ciencia, Innovación y Universidades, within projects MAT2015-73914- JIN, MAT2015-64083-R, and MAT2017-88358-C3-1-R, the Arag\'on Government project Q-MAD, and EU-QUANTERA project SUMO. We are grateful to Esteban Guti\'errez Mlot and Charles Downing for fruitful discussions.

\section*{References}

\bibliography{vortex}

\end{document}